\documentclass[12pt]{article}
\usepackage{graphicx}

\newcommand\pubdate{\today}

\def\indiana{Department of Physics\\
Indiana University \\
Bloomington, Indiana 47405-7105}
\def\Title#1{\begin{center} {\Large #1 } \end{center}}
\def\Author#1{\begin{center}{ \sc #1} \end{center}}
\def\Address#1{\begin{center}{ \it #1} \end{center}}

\newcommand\pubblock{\rightline{\begin{tabular}{l} \\
         \pubdate  \end{tabular}}}
\newenvironment{Abstract}{\begin{quotation}  }{\end{quotation}}
\newenvironment{Presented}{\begin{quotation} \begin{center} 
             PRESENTED AT\end{center}\bigskip 
      \begin{center}\begin{large}}{\end{large}\end{center} \end{quotation}}

\def\beq{\begin{equation}}
\def\eeq#1{\label{#1}\end{equation}}
\def\eeqn{\end{equation}}

\def\beqa{\begin{eqnarray}}
\def\eeqa#1{\label{#1}\end{eqnarray}}
\def\eeqan{\end{eqnarray}}

\let\bar=\overbar

\def\Dslash{\not{\hbox{\kern-4pt $D$}}}
\def\dslash{\not{\hbox{\kern-2pt $\del$}}}

\def\msb{{\bar{\ssstyle M \kern -1pt S}}}

\begin{document}
\begin{titlepage}
\pubblock

\vfill
\Title{A Review of $\chi_{cJ}(1P)$ Decays at BESIII and CLEO-c}
\vfill
%\Author{ Despina Reggiano\support}
\Author{ Ryan E. Mitchell}
\Address{\indiana}
\vfill
\begin{Abstract}
The latest results on $\chi_{cJ}(1P)$ decays from BESIII and CLEO-c are reviewed and compared to theoretical predictions.  The experimental results use the final samples of $\chi_{cJ}(1P)$ decays from CLEO-c, obtained from 26~million $\psi(2S)$ decays, and the most recent samples from BESIII, from a starting sample of 106~million $\psi(2S)$ decays.
\end{Abstract}
\vfill
\begin{Presented}
The 5th International Workshop on Charm Physics \\ (Charm 2012)\\
Honolulu, Hawai'i, USA,  May 14--17, 2012
\end{Presented}
\vfill
\end{titlepage}
\def\thefootnote{\fnsymbol{footnote}}
\setcounter{footnote}{0}

\section{Introduction}

\begin{figure}[b!]
\centering
\includegraphics[width=0.7\textwidth]{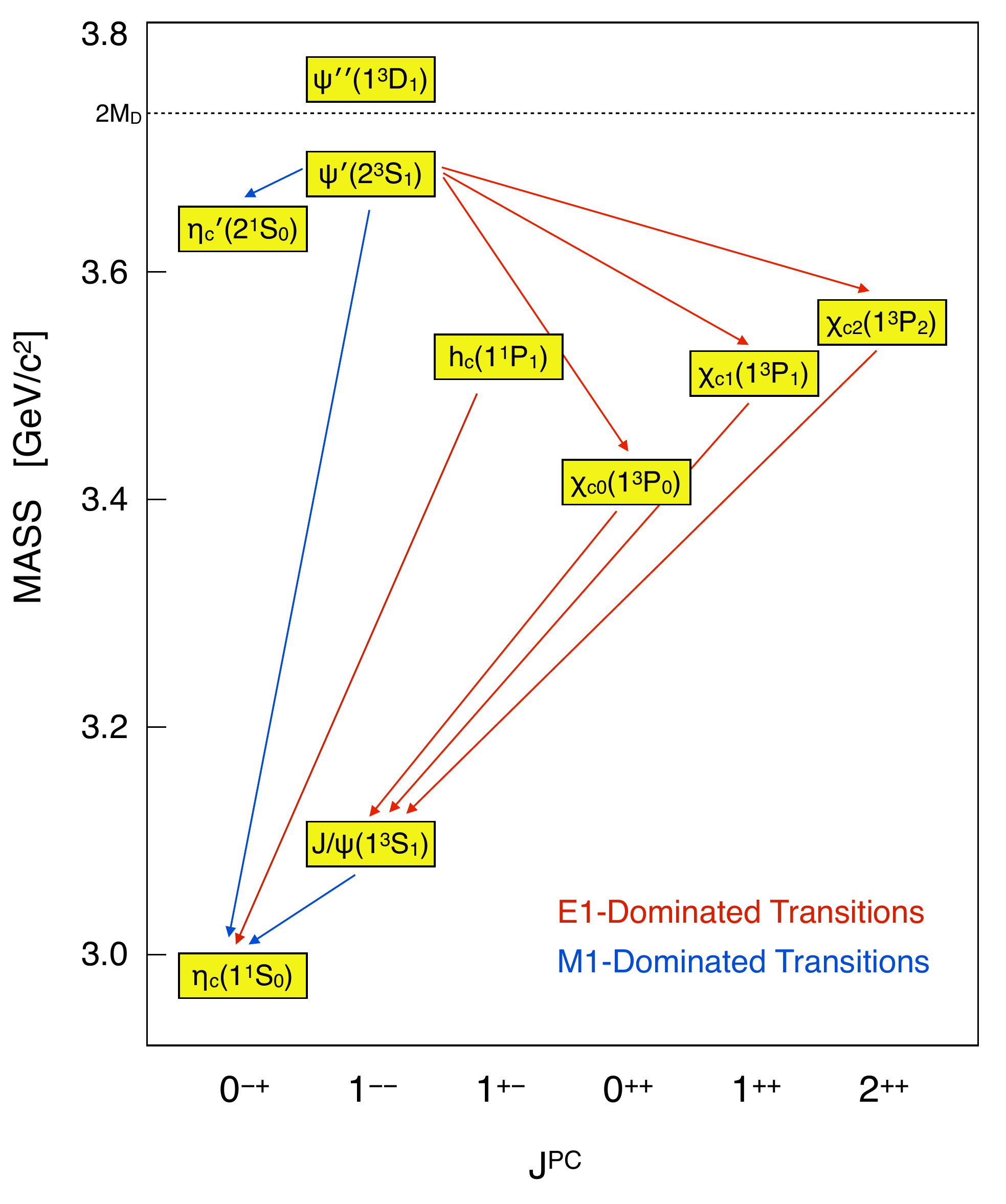}
\caption{The charmonium system.  The $\chi_{cJ}(1P)$ decays included in this review were obtained from $\psi(2S)$ radiative decays.}
\label{fig:charmonium}
\end{figure}

The $\chi_{cJ}(1P)$ states consist of a charm and an anticharm quark in a spin-1~(spin-aligned) state and with one unit of orbital angular momentum~($P$-wave) between them~(Fig.~\ref{fig:charmonium}). The total spin $J$ is thus 0, 1, or 2.  Since the $\chi_{cJ}$ are found abundantly in $\psi(2S)$ radiative decays, they can be most easily accessed at $e^+e^-$ colliders where the $\psi(2S)$ can be directly produced.  The $\chi_{cJ}$ decays covered in this review were collected by either the CLEO detector (starting with 26~million $\psi(2S)$ decays produced at the Cornell Electron Storage Ring~(CESR)) or BESIII (starting with 106~million $\psi(2S)$ decays produced at the Beijing Electron Positron Collider~(BEPCII)).

\begin{figure}[t!]
\centering
\includegraphics[width=0.9\textwidth]{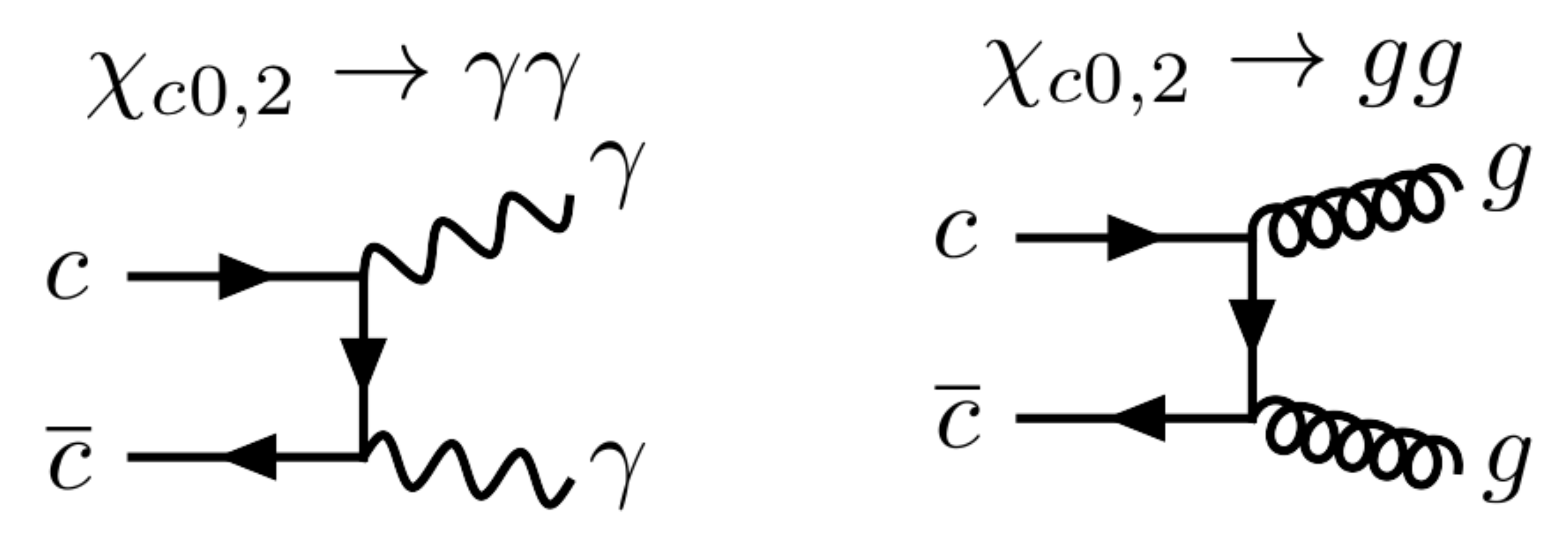}
\caption{The lowest order perturbative QCD diagrams for electromagnetic~(left) and strong~(right) decays of the $\chi_{c0}$ and $\chi_{c2}$.}
\label{fig:diagram1}
\end{figure}

Decays of the $\chi_{cJ}$ can be used as important probes of strong force dynamics.  The mass is large enough that the decays can be treated perturbatively and the lowest order diagrams are quite simple~(Fig.~\ref{fig:diagram1}), consisting of only charm-anticharm quark annihilation into two photons (for the lowest order electromagnetic process) or into two gluons (for the lowest order strong process).

\begin{figure}[htb]
\centering
\includegraphics[width=0.9\textwidth]{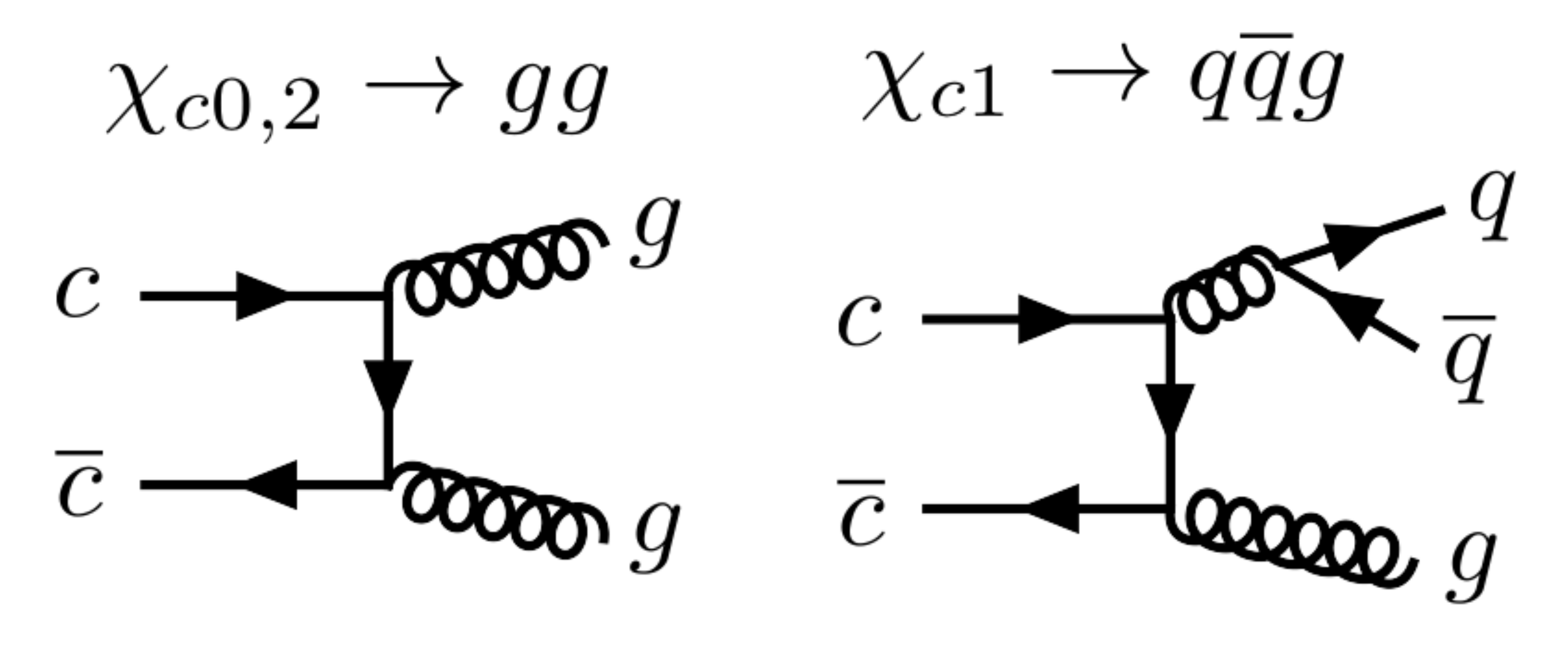} \\
\includegraphics[width=0.45\textwidth]{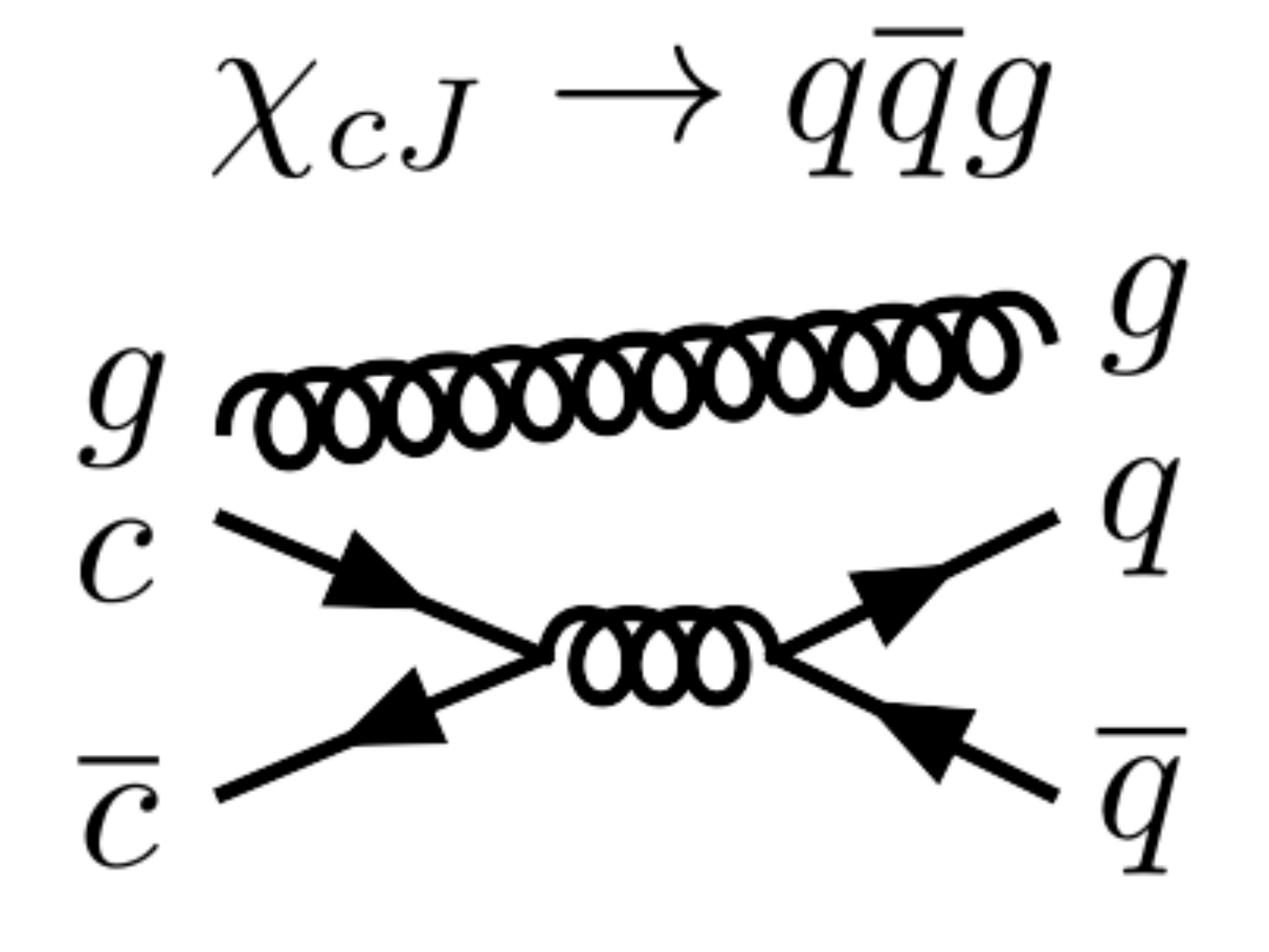}
\caption{The lowest order perturbative QCD diagrams for strong decays of the $\chi_{cJ}$.  The bottom figure assumes the initial $\chi_{cJ}$ is in a color octet state.}
\label{fig:diagram2}
\end{figure}

The lowest order diagram for $\chi_{c0,2}\to\gamma\gamma$ is pure QED, but the process is sensitive to QCD corrections.  Since several theoretical uncertainties cancel in the ratio of $\chi_{c2}$ to $\chi_{c0}$ two-photon widths, the ratio ${\cal R}$, defined as:
\begin{equation}
\label{eqn:r}
{\cal R} = \frac{\Gamma(\chi_{c2}\to\gamma\gamma)}{\Gamma(\chi_{c0}\to\gamma\gamma)},
\end{equation}
is particularly important.  Section~\ref{sec:gammagamma} will cover a new BESIII result for ${\cal R}$ and the two-photon widths of the $\chi_{c0}$ and $\chi_{c2}$.

The lowest order perturbative QCD diagrams for strong decays of the $\chi_{cJ}$ are shown in Fig.~\ref{fig:diagram2}.  The lowest order perturbative QCD diagrams do a poor job in many cases of predicting the patterns of $\chi_{cJ}$ decays.  This has led some to consider the possibility that the $\chi_{cJ}$ exists partially in a color octet state, where gluons play a role as constituent particles.  In this case, the simplest diagram is shown in the lower part of Fig.~\ref{fig:diagram2} and calculations have been performed using a Color Octet Model~(COM)~\cite{Wong:1999hc}.  Sections~\ref{sec:meson} and~\ref{sec:baryons} will review recent BESIII results for strong decays of the $\chi_{cJ}$.

In addition to decay dynamics, exclusive $\chi_{cJ}$ decays are also a good source of light quark states, which is useful for both light quark meson and baryon spectroscopy.  The rich set of final states available in $\chi_{cJ}$ decays allows one to isolate quantum numbers.  Section~\ref{sec:exotic} will review a recent analysis from CLEO, where an amplitude analysis was performed on the decays $\chi_{c1}\to\eta^{(\prime)}\pi^+\pi^-$ and evidence was found for a light quark state with exotic quantum numbers decaying to $\eta^{\prime}\pi$.

While CLEO concluded data-taking in 2008, the BESIII experiment continues to collect additional data in the charmonium region.  We thus expect further improvements in our understanding of $\chi_{cJ}$ decays in the near future.

\section{Electromagnetic Decays}
\label{sec:gammagamma}

As mentioned in the introduction, the lowest-order diagrams for $\chi_{c0,2}\to\gamma\gamma$ are purely QED, but the process is sensitive to higher-order QCD effects, such as radiative or relativistic corrections.  To lowest order in QED, the ratio of two-photon widths of the $\chi_{c0}$ and $\chi_{c2}$, ${\cal R}$, defined in Eqn.~\ref{eqn:r}, is $4/15\approx0.27$.  With QCD corrections, the predicted value ranges from 0.09 to 0.36 depending on the model~\cite{Gupta:1996ak,Godfrey:1985xj}.  Thus, precision measurements of ${\cal R}$ are important for arbitrating between models.

\begin{figure}[h!]
\centering
\includegraphics[width=0.9\textwidth]{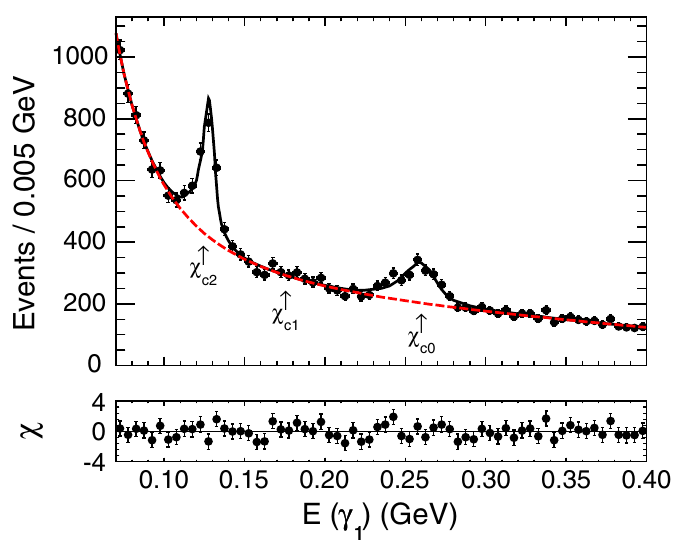} \\
\includegraphics[width=0.9\textwidth]{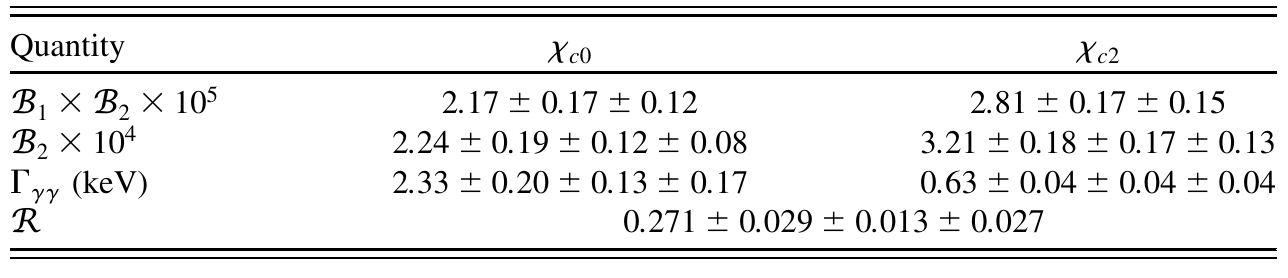}
\caption{Precision measurements of ${\cal B}_1(\psi(2S)\to\gamma\chi_{c0,2}) \times {\cal B}_2(\chi_{c0,2}\to\gamma\gamma)$ from BESIII~\cite{Ablikim:2012xi}.  The figures show the energy distribution of the photon from the $\psi(2S)$ to $\chi_{c0,2}$ transition and the $\chi$ distribution resulting from the fit.  The table shows the results, where $\Gamma_{\gamma\gamma}$ is the two-photon widths of the $\chi_{c0,2}$ and ${\cal R}$ is the ratio of the widths.  The first errors are statistical, the second systematic, and the third from PDG inputs (for ${\cal B}_1$ and the full widths of the $\chi_{c0}$ and $\chi_{c2}$). }
\label{fig:gammagamma}
\end{figure}

BESIII recently published new measurements of $\chi_{c0}$ and $\chi_{c2}$ two-photon widths~(and thus ${\cal R}$)~\cite{Ablikim:2012xi}, using the 3-photon process $\psi(2S)\to\gamma\chi_{c0,2}; \chi_{c0,2}\to\gamma\gamma$.  Results are shown in Fig.~\ref{fig:gammagamma}. The energy of the lowest-energy photon (called $\gamma_1$ in Fig.~\ref{fig:gammagamma}) was used to tag the $\chi_{c0}$ or $\chi_{c2}$.  The shape of the background, which is dominated by non-$\psi(2S)$ QED processes, was derived from data taken off-resonance at 3650~MeV.  The signal shapes were taken from a control sample of $\psi(2S)\to\gamma\chi_{c0,2}; \chi_{c0,2}\to K^+K^-$.  

\newpage
\newpage

The final results are also listed in Fig.~\ref{fig:gammagamma}, where ${\cal B}_1\equiv {\cal B}(\psi(2S)\to\gamma\chi_{c0,2})$ and ${\cal B}_2\equiv {\cal B}(\chi_{c0,2}\to\gamma\gamma)$.  For normalization, ${\cal B}_1$ and the full widths of the $\chi_{c0}$ and $\chi_{c2}$ were taken from the PDG~\cite{Nakamura:2010zzi}.  The final result for ${\cal R}$, perhaps surprisingly, is consistent with the lowest-order QED calculation.

%\begin{figure}[htb]
%\centering
%\includegraphics[width=0.7\textwidth]{figures/gammagamma_angles} 
%\caption{}
%\label{fig:charmonium}
%\end{figure}

\section{Strong Decays to Mesons}
\label{sec:meson}

\begin{figure}[htb]
\centering
\includegraphics[width=0.9\textwidth]{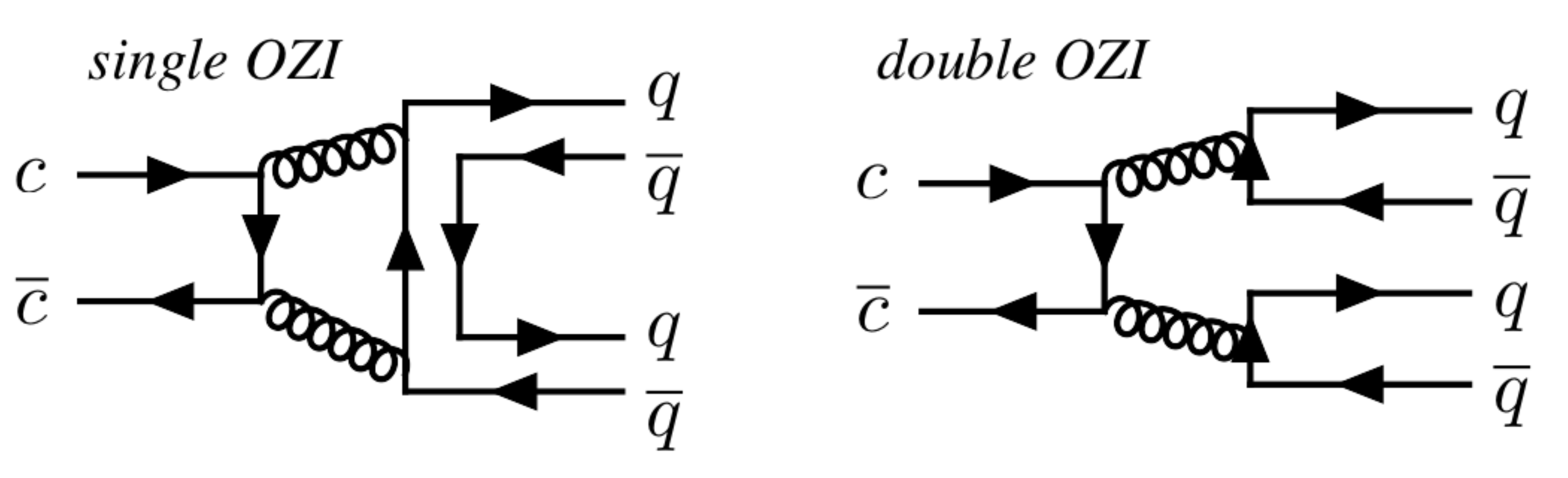} 
\caption{Diagrams for single and double OZI-violating decays of the $\chi_{cJ}$ to two vector states~($\omega,\phi$).  The lowest order diagrams for the decays $\chi_{cJ}\to\omega\phi$ are double OZI-violating. }
\label{fig:diagram4}
\end{figure}

The decays of the $\chi_{cJ}$ into two vector mesons ($J^{PC}=1^{--}$), for example $\chi_{cJ} \to \omega\omega,\omega\phi,\phi\phi$, have a number of interesting features.  First, from perturbative QCD, one would expect the branching fractions to be smaller than $10^{-3}$.  Second, the helicity selection rule predicts the decays of the $\chi_{c1}$ to two vector mesons to be suppressed.  And finally, the decays $\chi_{cJ}\to\phi\phi,\omega\omega$ are only singly OZI-suppressed (as are all charmonium decays), while the decays $\chi_{cJ}\to\omega\phi$ are doubly OZI-suppressed.  This can be seen in the diagrams shown in Fig.~\ref{fig:diagram4}.

\begin{figure}[h!]
\centering
\includegraphics[width=0.75\textwidth]{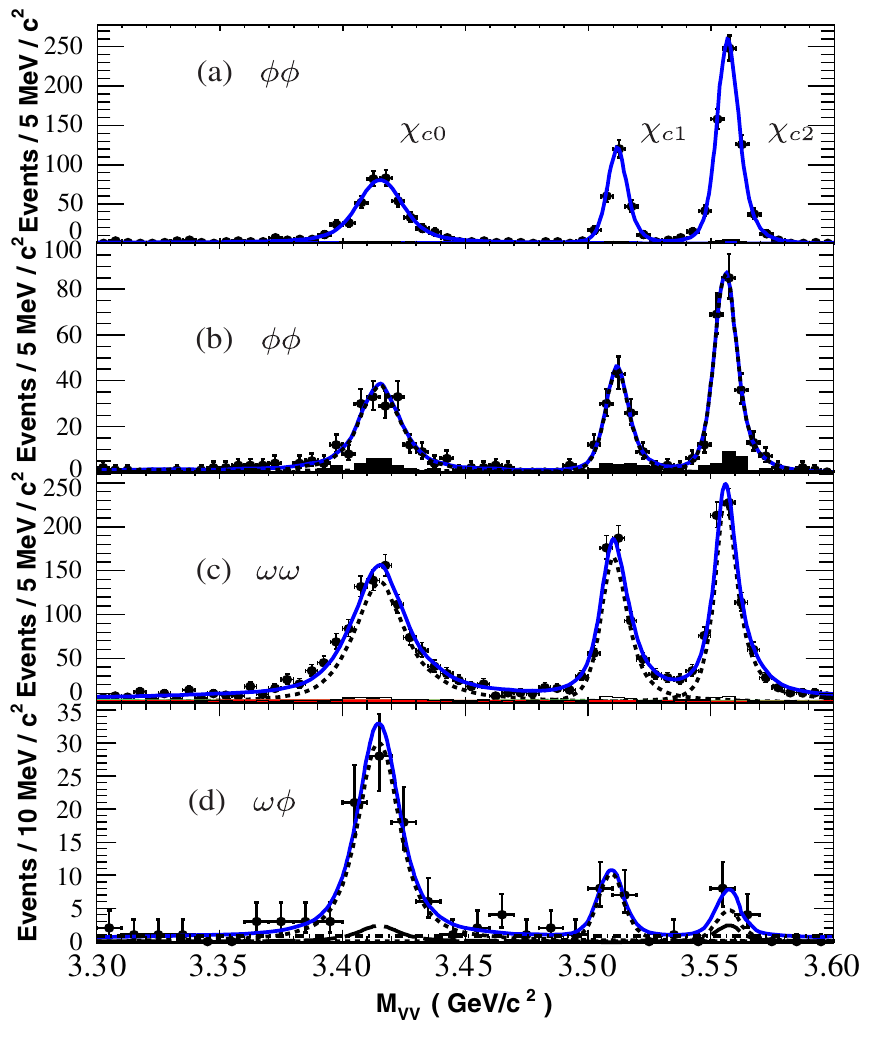} 
\caption{Measurements of $\chi_{cJ}$ decays to (a,b)~$\phi\phi$, (c)~$\omega\omega$, and (d)~$\omega\phi$ from BESIII~\cite{Ablikim:2011aa}.  In (b), one of the $\phi$ decays to $\pi^+\pi^-\pi^0$.  In all other cases the $\phi$ decays to $K^+K^-$.}
\label{fig:vv}
\end{figure}

\begin{figure}[h!]
\centering
\includegraphics[width=0.8\textwidth]{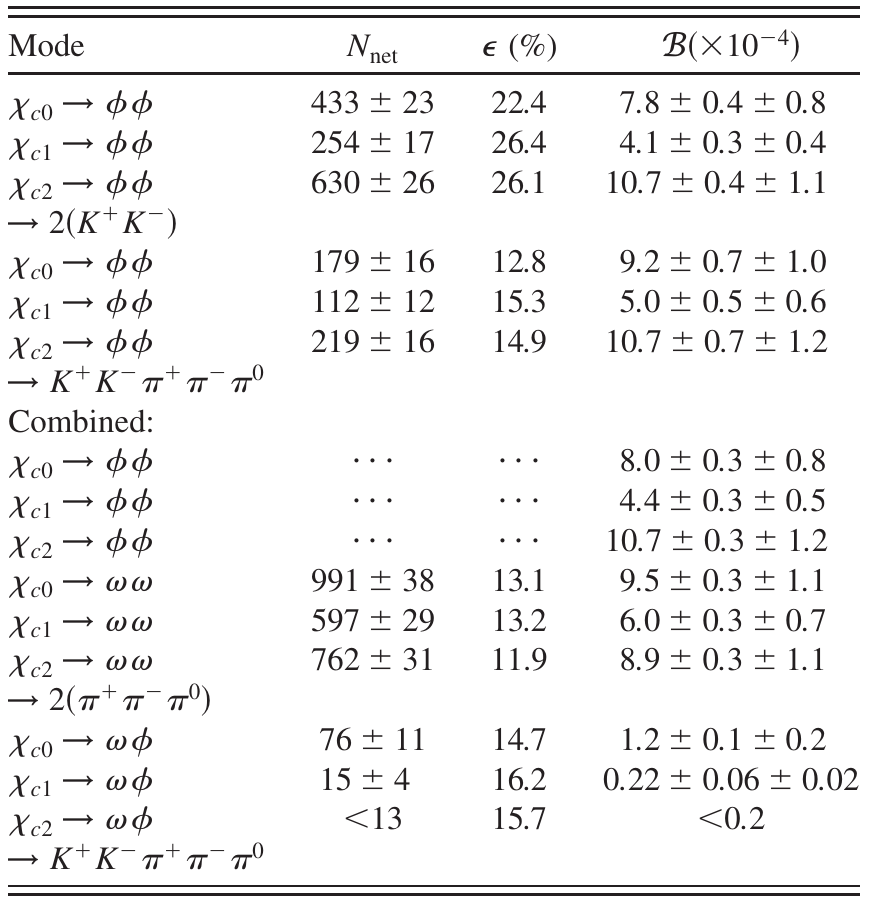} 
\caption{The number of observed events~($N_{\mathrm{net}}$), the efficiencies~($\epsilon$), and the branching fractions~(${\cal B}$) for the decay modes listed in the first column.  The results are from BESIII~\cite{Ablikim:2011aa}. }
\label{fig:vvtable}
\end{figure}

BESIII published new results on $\chi_{cJ}\to\omega\omega,\phi\phi,\omega\phi$ in 2011~\cite{Ablikim:2011aa} that address these issues.  The results are shown in Fig.~\ref{fig:vv}, where the small peaking backgrounds have been estimated from $\omega$ or $\phi$ sidebands.  First, the branching fractions are on the order of $10^{-3}$, somewhat higher than one would expect from perturbative QCD.  Second, there are substantial rates for the $\chi_{c1}$ decays, which appears to contradict expectations from the helicity selection rule.  Finally, BESIII made the first observations of the doubly OZI-suppressed $\chi_{cJ}\to\omega\phi$ decays, with branching fractions roughly an order of magnitude down from the singly OZI-suppressed decays.  The resulting numbers are shown in Fig.~\ref{fig:vvtable}.

\newpage
\newpage

\newpage

\section{Strong Decays to Baryons}
\label{sec:baryons}

The simplest baryon decays of the $\chi_{cJ}$, namely $\chi_{c1}\to p\overline{p}$ and $\chi_{c2}\to p\overline{p}$, have long presented a theoretical challenge.  While the perturbative QCD calculations, assuming the $\chi_{cJ}$ is in a color singlet state, predict ${\cal B}(\chi_{c1}\to p\overline{p}) = 0.29\times 10^{-5}$ and ${\cal B}(\chi_{c2}\to p\overline{p}) = 0.84\times 10^{-5}$, the experimental measurements are an order of magnitude larger.  Furthermore, the decay $\chi_{c0}\to p\overline{p}$ (and $\chi_{c0}$ decays to baryon anti-baryon pairs in general) should be suppressed in perturbative QCD by the ``helicity selection rule,'' which links the final state hadrons to the helicities of the intermediate gluons.  But experimentally, substantial $\chi_{c0}$ decay rates have been found.

\begin{figure}[b!]
\centering
\includegraphics[width=1.0\textwidth]{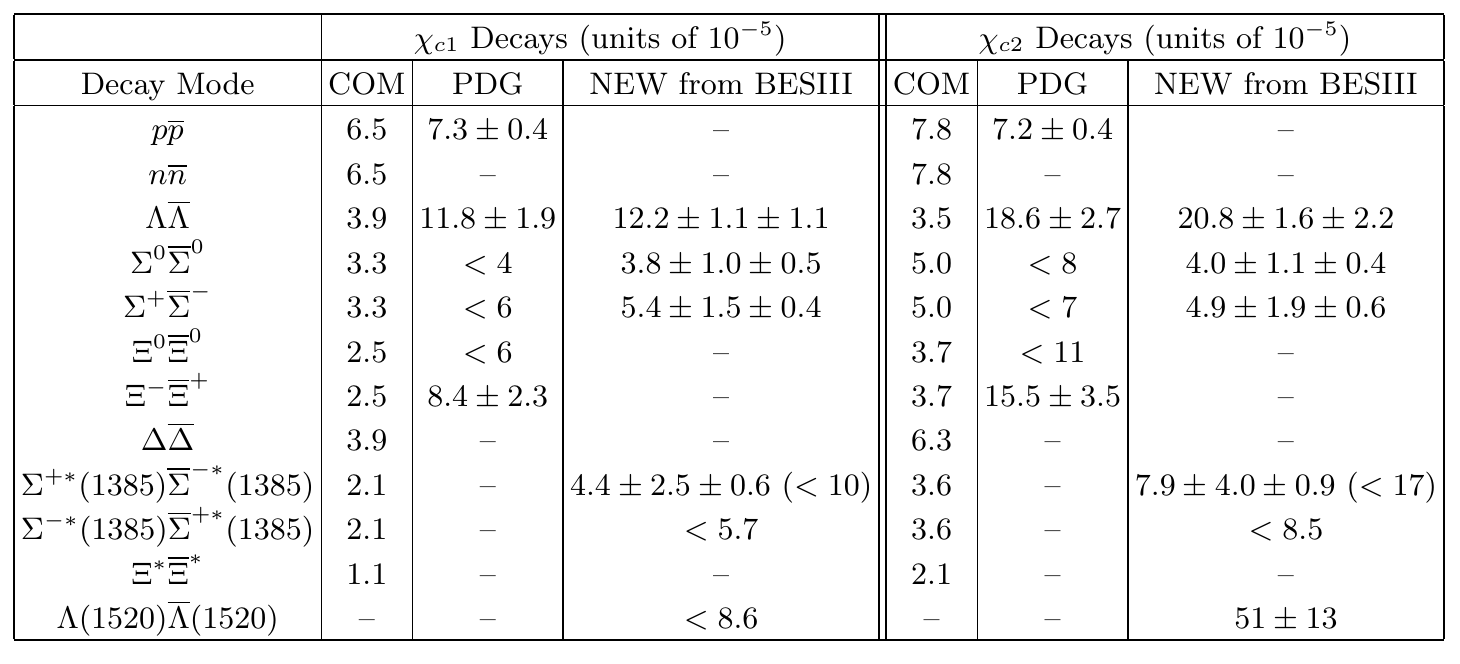} 
\caption{The Color Octet Model~(COM)~\cite{Wong:1999hc} predictions, the Particle Data Group~(PDG)~\cite{Nakamura:2010zzi} values, and new BESIII measurements~\cite{:2012hi,Ablikim:2011uf} of branching fractions of $\chi_{c1}$ and $\chi_{c2}$ decays to various di-baryon final states.  All branching fractions are in units of $10^{-5}$.}
\label{fig:com}
\end{figure}

To address the discrepancy between theory and experiment in $\chi_{c1}$ and $\chi_{c2}$ $p \overline{p}$ decays, a ``color octet model''~(COM) was devised, which allows the $\chi_{cJ}$ to have a color octet component in which there is a valence gluon~\cite{Wong:1999hc}.  This model was able to bring the theoretical $p\overline{p}$ branching fractions into alignment with experiment, but still underestimated other baryon-antibaryon decays such as $\Lambda\overline{\Lambda}$.  A comparison of COM predictions and experimental measurements from the PDG are shown in Fig.~\ref{fig:com}.

Inspired by these COM predictions, there has recently been a series of analyses from BESIII looking at $\chi_{cJ}$ decays to baryon pairs~\cite{:2012hi,Ablikim:2011uf}.  The new results from BESIII are also listed in Fig.~\ref{fig:com}.

\begin{figure}[t!]
\centering
\includegraphics[width=0.9\textwidth]{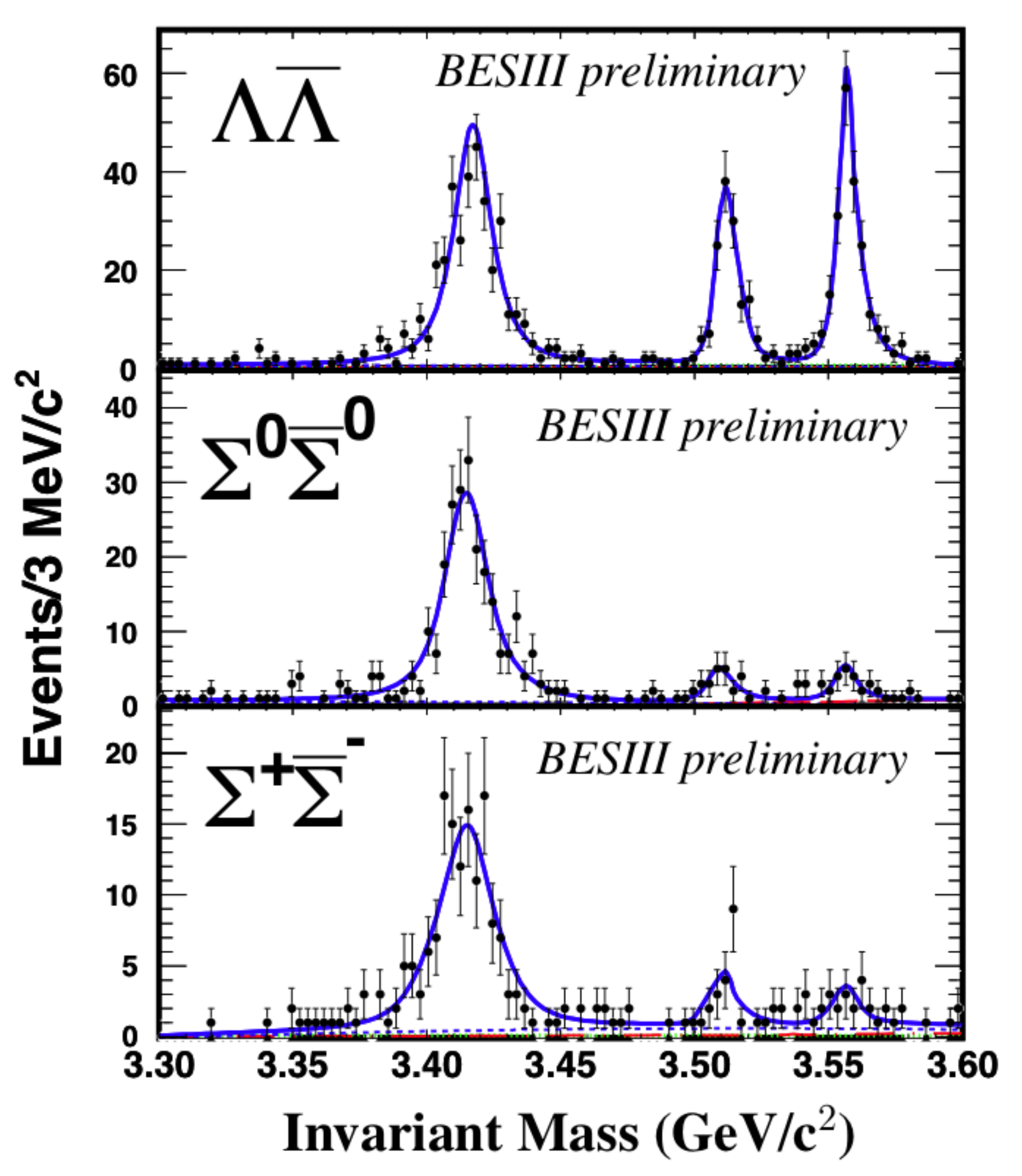}
\caption{Preliminary measurements of branching fractions of $\chi_{cJ}$ decays to $\Lambda\overline{\Lambda}$, $\Sigma^0\overline{\Sigma}^0$, and $\Sigma^+\overline{\Sigma}^-$ from BESIII.}
\label{fig:bb1}
\end{figure}

The first BESIII measurements are of $\chi_{cJ}$ decays to $\Lambda\overline{\Lambda}$, $\Sigma^0\overline{\Sigma}^0$, and $\Sigma^+\overline{\Sigma}^-$.  The signals are shown in Fig.~\ref{fig:bb1} and the $\chi_{c1}$ and $\chi_{c2}$ branching fractions are listed in Fig.~\ref{fig:com}.  The $\Lambda\overline{\Lambda}$ branching fractions are still significantly larger than the COM predictions, while the $\Sigma\overline{\Sigma}$ decays appear to be consistent, although with large statistical errors.  Interestingly, the $\chi_{c0}$ branching fractions, which are predicted to be suppressed by the helicity selection rule, are far larger than those of the $\chi_{c1,2}$.  The preliminary results from BESIII for the $\chi_{c0}$ decays are ${\cal B}(\chi_{c0}\to\Lambda\overline{\Lambda}) = (33.3\pm2.0\pm2.6)\times10^{-5}$, ${\cal B}(\chi_{c0}\to\Sigma^0\overline{\Sigma}^0) = (47.8\pm3.4\pm3.8)\times10^{-5}$, and ${\cal B}(\chi_{c0}\to\Sigma^+\overline{\Sigma}^-) = (45.4\pm4.2\pm2.5)\times10^{-5}$.

\begin{figure}[t!]
\centering
\includegraphics[width=0.9\textwidth]{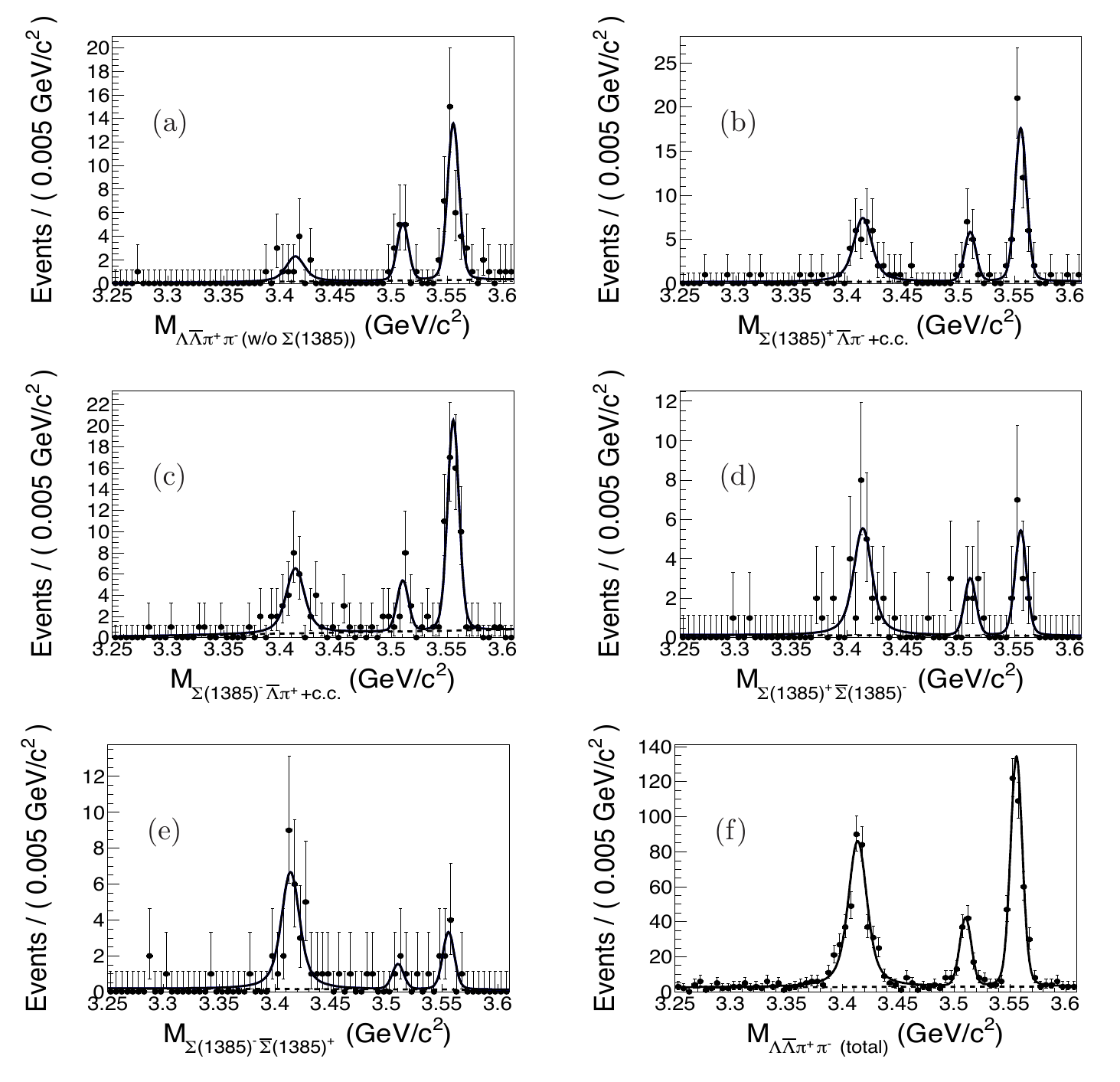}
\caption{BESIII measurements of $\chi_{cJ}$ yields in decays to $\Lambda\overline{\Lambda}\pi^+\pi^-$ in regions corresponding to (a)~no $\Sigma(1385)$, (b)~the $\Sigma(1385)^+$, (c)~the $\Sigma(1385)^-$, (d)~the $\Sigma(1385)^+$ and $\overline{\Sigma}(1385)^-$, (e)~the $\Sigma(1385)^-$ and $\overline{\Sigma}(1385)^+$, and (f)~all regions~\cite{:2012hi}. }
\label{fig:bb2}
\end{figure}

\begin{figure}[t!]
\centering
\includegraphics[width=1.0\textwidth]{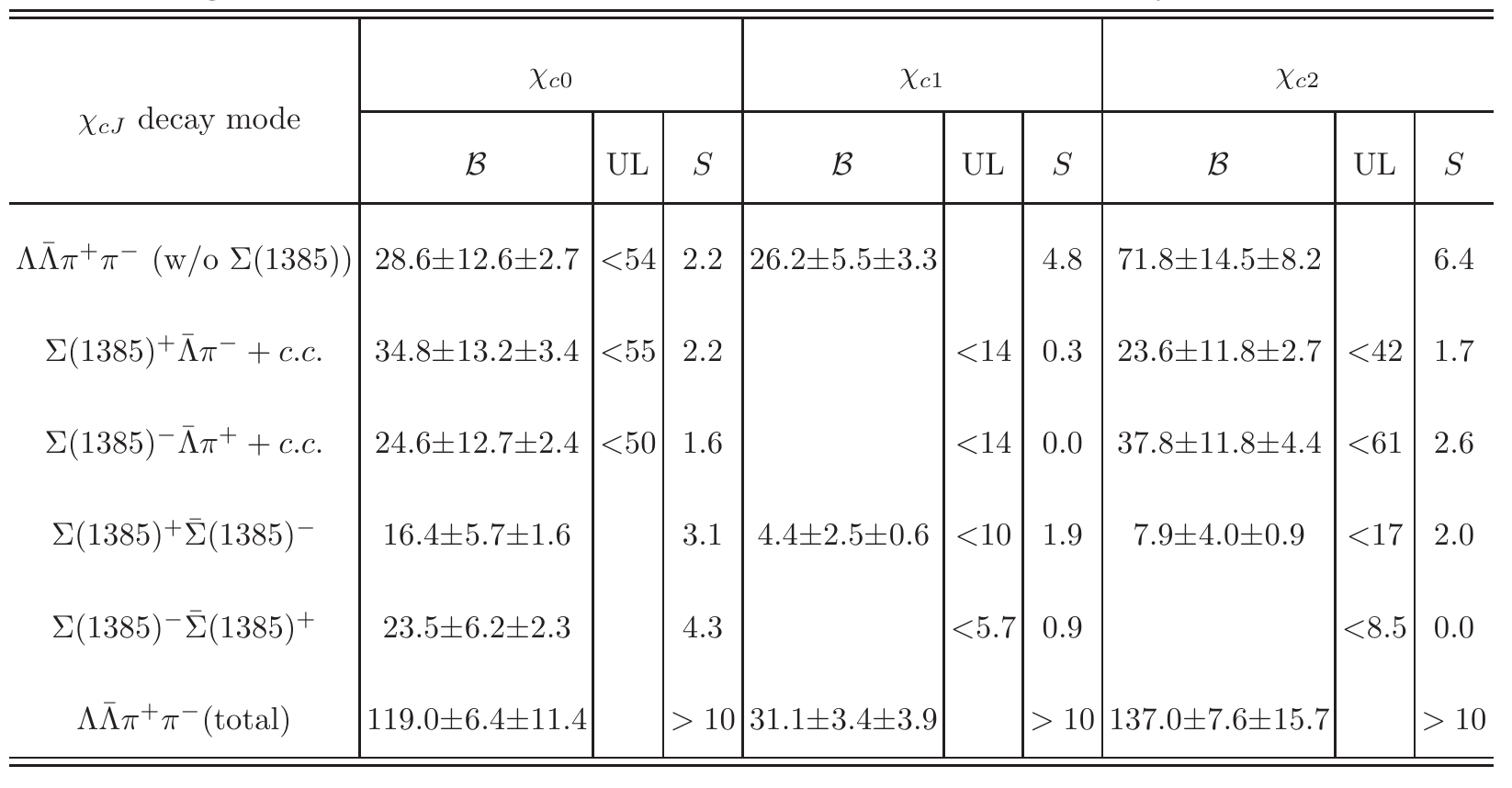} 
\caption{BESIII measurements of $\chi_{cJ}$ branching fractions to $\Lambda\overline{\Lambda}\pi^+\pi^-$ and various combinations of internal $\Sigma(1385)$ decays~\cite{:2012hi}.  Branching fractions~(${\cal B}$) are in units of $10^{-5}$ and upper limits~(UL) are at 90\% confidence level.  The significance of each channel is given in the $S$ column.}
\label{fig:bb2table}
\end{figure}

The second BESIII measurements~\cite{:2012hi} consist of $\chi_{cJ}$ decays to $\Lambda\overline{\Lambda}\pi^+\pi^-$ and all of the intermediate processes that can be produced with the $\Sigma(1385)$.  Of particular interest for comparison with the COM are the $\chi_{c1,2}$ decays to $\Sigma(1385)^+\overline{\Sigma}(1385)^-$ and $\Sigma(1385)^-\overline{\Sigma}(1385)^+$. The $\chi_{cJ}$ peaks for various regions of $\Lambda\overline{\Lambda}\pi^+\pi^-$ sub-masses are shown in Fig.~\ref{fig:bb2}.  Since the $\Sigma(1385)$ is fairly wide, and overlaps significantly with $\Lambda\overline{\Lambda}\pi^+\pi^-$ phase space, the various branching fractions (with different combinations of $\Sigma(1385)$) were extracted using a matrix of efficiencies and overlaps determined from signal Monte Carlo.  The results are listed in Fig.~\ref{fig:bb2table}.  The $\chi_{c1}$ and $\chi_{c2}$ decays are consistent with the COM predictions (within large statistical errors).  But again, the $\chi_{c0}$ decays appear to be dominant.

\begin{figure}[h!]
\centering
\includegraphics[width=1.0\textwidth]{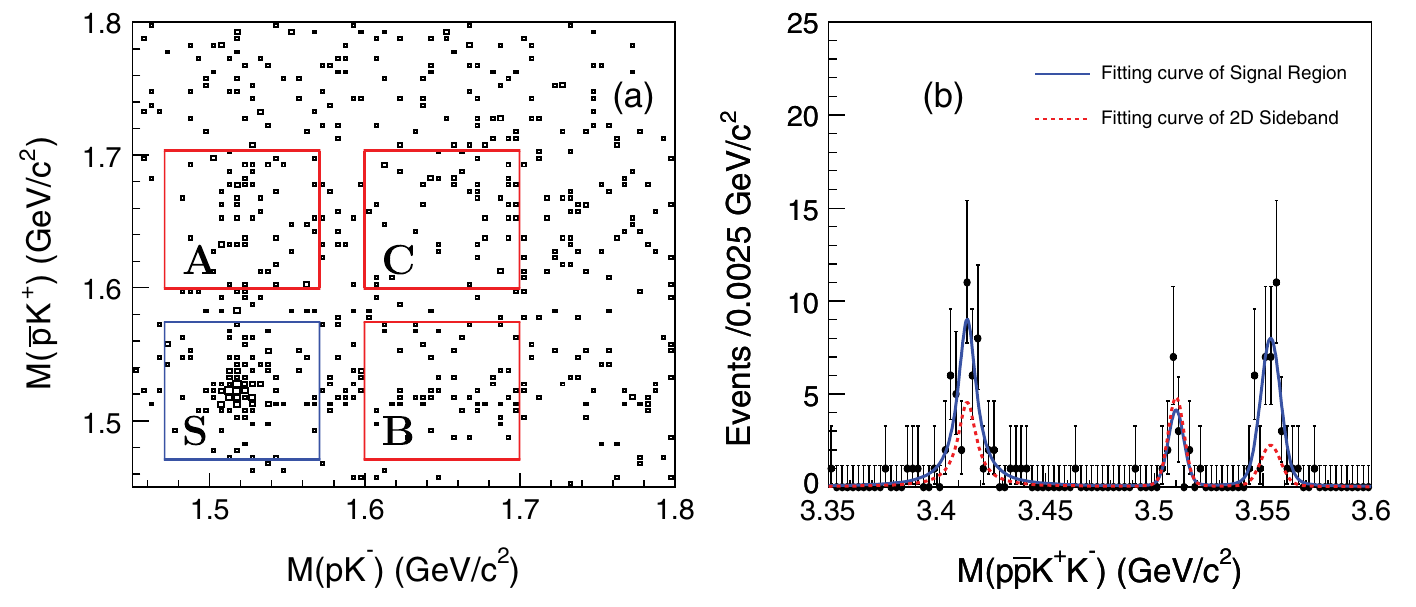}\\ 
\includegraphics[width=1.0\textwidth]{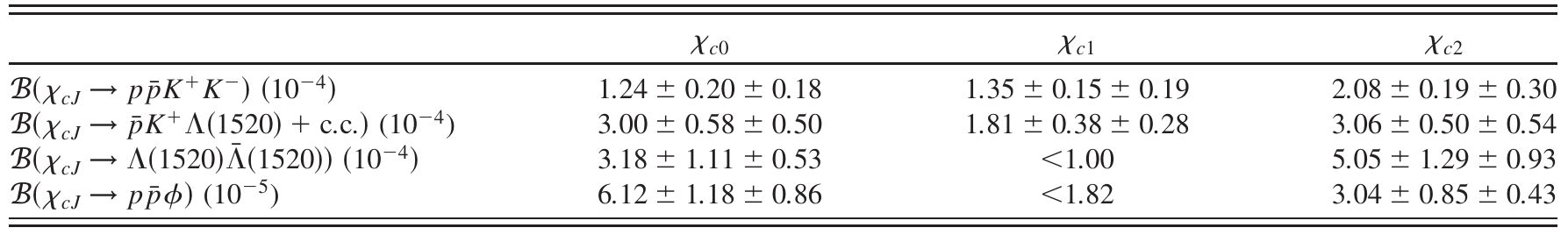} 
\caption{BESIII measurements of $\chi_{cJ}\to p\overline{p}K^+K^-$ and its submodes involving $\Lambda(1520)\to pK^-$, $\overline{\Lambda}(1520)\to\overline{p}K^+$, and $\phi\to K^+K^-$.  The upper plots show the selection of the $\Lambda(1520)\overline{\Lambda}(1520)$ signal region~(left) and the resulting $\chi_{cJ}$ peaks~(right).  The solid~(blue) line is a fit in the signal region and the dotted~(red) line is a fit in the sidebands.  The table lists the resulting branching fractions.}
\label{fig:bb3}
\end{figure}

Finally, BESIII has also measured the decays $\chi_{cJ}\to\Lambda(1520)\overline{\Lambda}(1520)$ using the final state $K^+K^-p\overline{p}$.  Signal and sideband regions were defined in $M(\overline{p}K^+)$ and $M(pK^-)$ and a simultaneous fit was performed to these regions, as shown in Fig.~\ref{fig:bb3}.  The final results are also listed in Fig.~\ref{fig:bb3}.  Perhaps surprisingly, the rates for $\chi_{c0}$ and $\chi_{c2}$ decays to $\Lambda(1520)\overline{\Lambda}(1520)$ are as large as those to $\Lambda\overline{\Lambda}$ (which were already larger than the COM predictions).

As can be seen in Fig~\ref{fig:com}, the COM is successful (within experimental uncertainties) in describing many $\chi_{c1}$ and $\chi_{c2}$ decays into baryons and anti-baryons.  The $\Lambda\overline{\Lambda}$ channel is the notable exception.  Since the BESIII measurements were inspired by the COM, the focus has been on comparisons between experiment and that model.  Note that other models exist, however, and can be found extensively reviewed in~\cite{Brambilla:2010cs}.

%\begin{figure}[htb]
%\centering
%\includegraphics[width=0.9\textwidth]{figures/BB_table}\\ 
%\caption{Preliminary measurements of branching fractions of $\chi_{cJ}$ decays to $\Lambda\overline{\Lambda}$, $\Sigma^0\overline{\Sigma}^0$, and $\Sigma^+\overline{\Sigma}^-$ from BESIII~(``This work'')~\cite{xxxx}.}
%\label{fig:charmonium}
%\end{figure}

\newpage

\newpage

\section{Light Meson Spectroscopy in $\chi_{cJ}$ Decays}
\label{sec:exotic}

\begin{figure}[h!]
\centering
\includegraphics[width=1.0\textwidth]{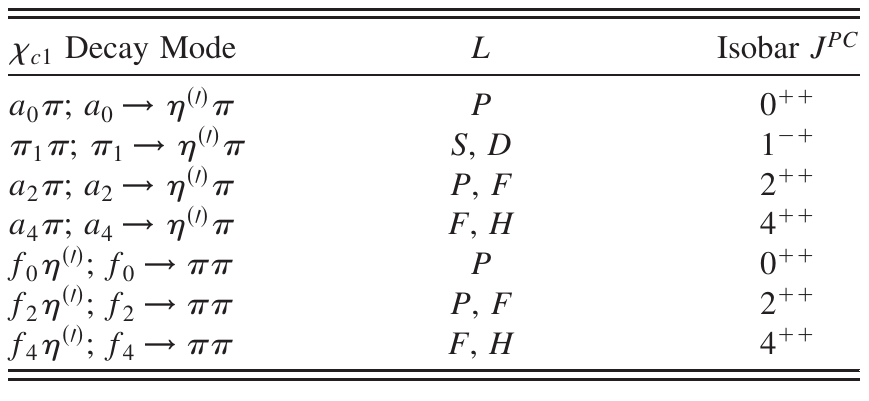} 
\caption{A list of allowed decays of the $\chi_{c1}$ that result in the final state $\eta^{(\prime)}\pi^+\pi^-$, the angular momentum~($L$) of the initial $\chi_{c1}$ decay, and the $J^{PC}$ of the intermediate state.  Notice that the decay through the exotic $\pi_1$ state is the only allowed $S$-wave decay and thus could be enhanced relative to other decay modes.}
\label{fig:cleo1}
\end{figure}

Decays of the $\chi_{cJ}$ are not only important for the study of the strong force through the dynamics of their decays, but they also serve as a source of light quark states.  The large number of available final states, and the fact the initial state can have $J=0,1,2$, depending on the $\chi_{cJ}$ state, allows one to optimize searches for light quark mesons with specific quantum numbers.  The decays $\chi_{c1}\to\eta\pi^+\pi^-$ and $\chi_{c1}\to\eta^{\prime}\pi^+\pi^-$, for example, are an ideal place to search for resonances with exotic $J^{PC}=1^{-+}$ quantum numbers decaying to $\eta\pi$ or $\eta^\prime\pi$.  The possible resonances produced in $\chi_{c1}\to\eta^{(\prime)}\pi^+\pi^-$ decays are listed in Fig.~\ref{fig:cleo1}.  Here the only allowed $S$-wave decay of the $\chi_{c1}$ goes through the exotic $\pi_1$ state, which in turn decays to $\eta^{(\prime)}\pi$.

\begin{figure}[htb]
\centering
\includegraphics[width=1.0\textwidth]{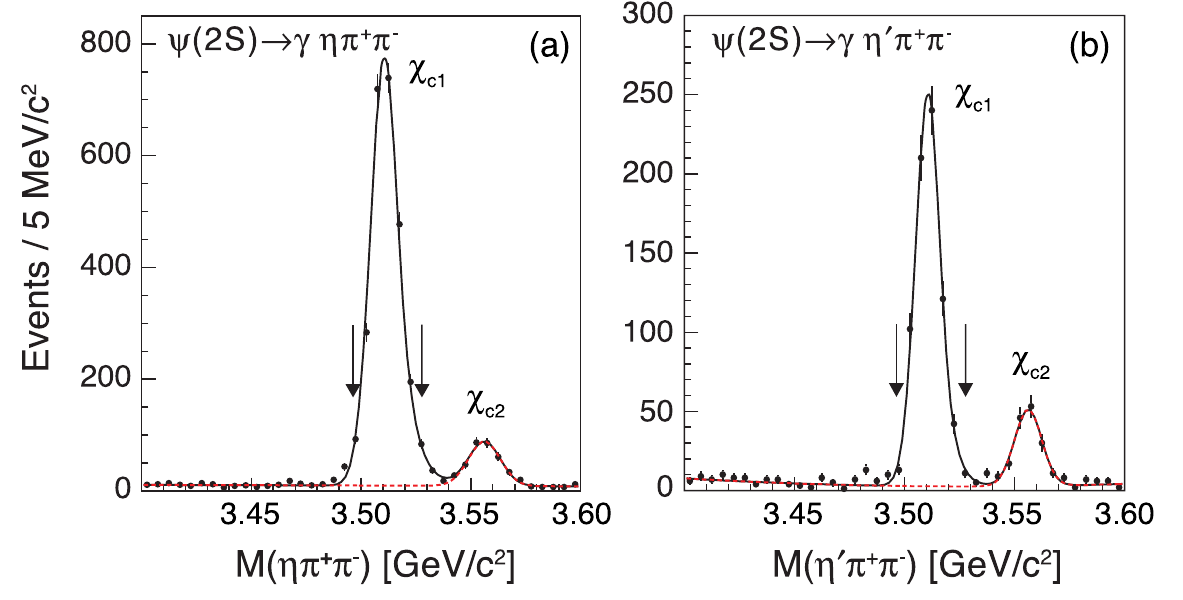} 
\caption{The selection of $\chi_{c1}$ decays to $\eta\pi^+\pi^-$~(left) and $\eta^\prime\pi^+\pi^-$~(right) from CLEO~\cite{Adams:2011sq}.  The signal purities are estimated to be 97.5\% and 94.6\%, respectively.}
\label{fig:cleo2}
\end{figure}

\begin{figure}[h!]
\centering
\includegraphics[width=1.0\textwidth]{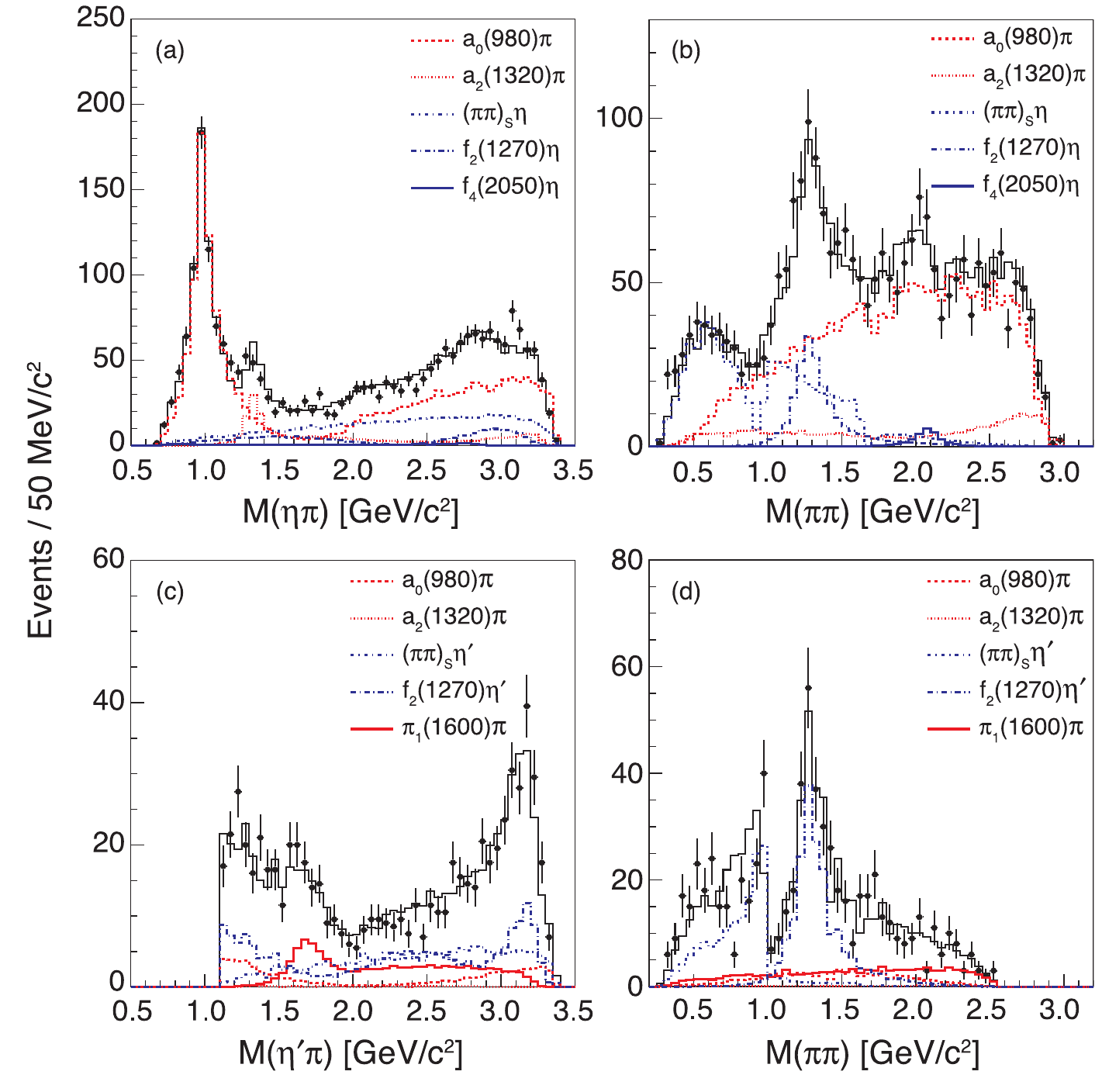} 
\caption{The fit results for $\chi_{c1}\to\eta\pi^+\pi^-$~(top) and $\chi_{c1}\to\eta^\prime\pi^+\pi^-$~(bottom) from CLEO~\cite{Adams:2011sq}.  The points are data and the solid black line is the total fit result.  The other lines show contributions from individual amplitudes.  The exotic $\pi_1$ signal decaying to $\eta^\prime\pi$ can be seen in the lower left.}
\label{fig:cleo3}
\end{figure}

\begin{figure}[htb]
\centering
\includegraphics[width=1.0\textwidth]{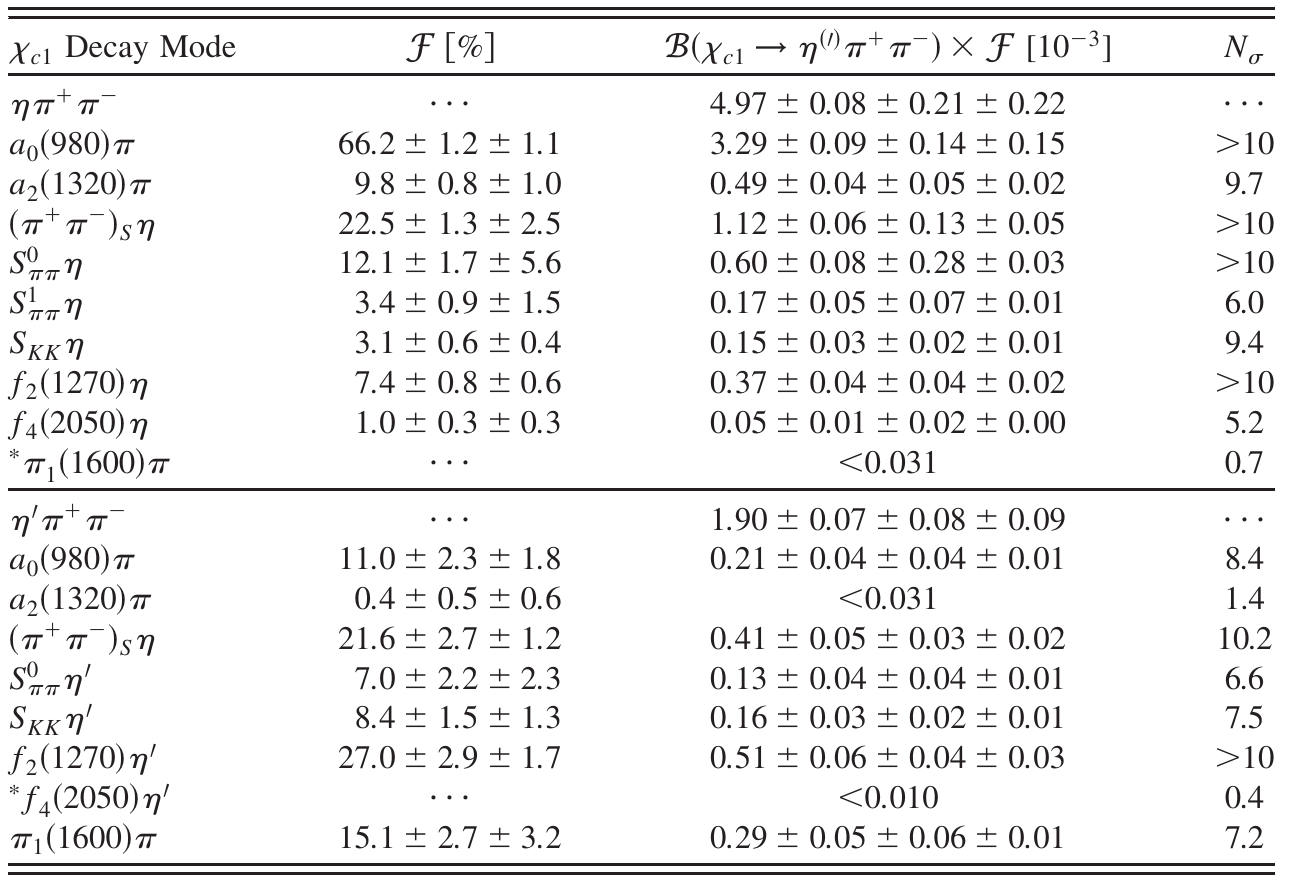} 
\caption{Results from the CLEO amplitude analyses of $\chi_{c1}\to\eta^{(\prime)}\pi^+\pi^-$ showing the fit fractions~(${\cal F}$) of different submodes, the branching fraction of $\chi_{c1}\to\eta^{(\prime)}\pi^+\pi^-$ times the fit fractions~(${\cal B}\times{\cal F}$), and the significance of each fit component~($N_{\sigma}$)~\cite{Adams:2011sq}.  The first errors are statistical, the second systematic, and the third, where applicable, are from the external measurement of ${\cal B}(\psi(2S)\to\gamma\chi_{c1})$.}
\label{fig:cleo4}
\end{figure}

The CLEO Collaboration has recently published an amplitude analysis of the decays $\chi_{c1}\to\eta^{(\prime)}\pi^+\pi^-$~\cite{Adams:2011sq}.  Clean samples of the $\chi_{c1}$, shown in Fig.~\ref{fig:cleo2}, were obtained from 26~million $\psi(2S)$ decays.  An amplitude analysis was then performed using the amplitudes listed in Fig.~\ref{fig:cleo1}.  Breit-Wigner distributions were used to parameterize most resonance decays, while a Flatte distribution was used for $a_0(980)$ decays, and a phenomenological distribution taken from $\pi\pi$ scattering data was used for the $\pi\pi$ $S$-wave (called $f_0$ in Fig.~\ref{fig:cleo1}).  Projections of the fit results are shown in Fig.~\ref{fig:cleo3} and the results are listed in Fig.~\ref{fig:cleo4}.

In the $\chi_{c1}\to\eta^\prime\pi^+\pi^-$ channel, CLEO found evidence for an exotic $\pi_1$ state decaying to $\eta^\prime\pi$, which is consistent with previous claims of a $\pi_1(1600)$ state produced in other production mechanisms~\cite{Beringer:1900zz}.  This is the first evidence of a light quark meson with exotic quantum numbers in a charmonium decay, and opens many new possibilities for other amplitude analyses of $\chi_{cJ}$ decays into other final states.

\newpage

\newpage

\newpage

\section{Conclusion}

While data taking has concluded at CLEO, and no new $\chi_{cJ}$ analyses are anticipated, BESIII continues to collect data in the charmonium region.  The BESIII results presented here are based on 106~million $\psi(2S)$ decays, but more statistics are expected.  With larger statistics, we expect to learn more about strong decay dynamics and perform more searches for light quark states in $\chi_{cJ}$ decays.

%\Acknowledgements

\end{document}